# Relativistic derivations of the electric and magnetic fields generated by an electric point charge moving with constant velocity

Bernhard Rothenstein[1], Stefan Popescu[2] and George J. Spix[3]

1) Politehnica University of Timisoara, Physics Department, Timisoara, Romania, bernhard_rothenstein@yahoo.com
2) Siemens AG, Erlangen, Germany, stefan.popescu@siemens.com
3) BSEE Illinois Institute of Technology, USA, gjspix@msn.com

**Abstract**. *We propose a simple relativistic derivation of the electric and the magnetic fields generated by an electric point charge moving with constant velocity. Our approach is based on the radar detection of the point space coordinates where the fields are measured. The same equations were previously derived in a relatively complicated way² based exclusively on general electromagnetic field equations and without making use of retarded potentials or relativistic equations.*

## 1. Introduction

Heaviside[1] obtained the expressions for the electric and magnetic fields produced by an electric point charge $q$ moving with constant velocity $V$ given by

$$\mathbf{E} = \frac{q}{4\pi\varepsilon_0 r^3} \frac{1-\beta^2}{\left[1-\beta^2 \sin^2\theta\right]^{3/2}} \mathbf{r} \qquad (1)$$

$$\mathbf{B} = c^{-2}\mathbf{V} \times \mathbf{E} \qquad (2)$$

where $r$ is the distance between the point of observation and the charge, $c$ is the velocity of light and $\theta$ is the angle between $\mathbf{r}$ and $\mathbf{V}$.

Jefimenko[2] presents the ways in which (1) and (2) could be derived. Among the four different approaches that he mentions, one particular approach is based on applying Lorentz-Einstein transformations of space-time coordinates and fields to a stationary point charge.[3] He considers that this method is fairly simple mathematically, but it doesn't fit into the classical theory of electromagnetism.

In the literature covering this subject we found the following formulas

$$\mathbf{E} = \frac{q}{4\pi\varepsilon_0 r^3} \frac{1-\beta^2}{\left(1-\frac{\mathbf{r}.\mathbf{v}}{rc}\right)^3} \left[\mathbf{r} - \frac{r\mathbf{V}}{c}\right] \qquad (3)$$

$$\mathbf{B} = \mu_0 \frac{q}{4\pi r^3} \frac{1-\beta^2}{\left[1-\frac{\mathbf{r}.\mathbf{V}}{rc}\right]^3} [\mathbf{V} \times \mathbf{r}]. \qquad (4)$$

Jefimenko[2] derives them based exclusively on general electromagnetic field equations and do not make use of retarded potentials or relativistic equations. These derivations are relatively complicated making hard theirs teaching without mnemonic helps.

The purpose of our paper is to present the relativistic derivations of (1) and (2) and of (3) and (4) in a simple and transparent way showing the difference between the physics behind them. We start with the hints that the relativists always have in mind:

• When you speak about a physical quantity then always define it without ambiguity mentioning the observer who measures it, the measuring devices he uses and where and when he performs this measurement.

**2. The electric and the magnetic fields of an electric point charge moving with constant velocity.**

The scenario we propose involves the inertial reference frames K(XOY) and K'(X'O'Y') with parallel axes, K' moving with constant speed $V$ relative to K in the positive direction of the common OX(O'X') axes. At a time $t=t'=0$ the origins O and O' coincide in space. A point charge $q$ is at rest in K' being located at its origin O' as shown in Figure 1. The electric field generated by this charge (stationary in K') is measured at a point $M'(x' = r'\cos\theta', y' = r'\sin\theta')$ by the observer $R'(x' = r'\cos\theta', y' = r'\sin\theta')$ located at this point. Assuming that the Coulomb's law holds true in the rest frame of the charge, then the point charge $q$ creates a radial electric field in K'

$$E' = \frac{q}{4\pi\varepsilon_0 r'^2} = \frac{kq}{r'^2} \tag{5}$$

The components of which are

$$E'_x = \frac{kq}{r'^2}\cos\theta' \tag{6}$$

and

$$E'_y = \frac{kq}{r'^2}\sin\theta'. \tag{7}$$

In its rest frame K' the charge doesn't' generate a magnetic field thus $B'=0$.

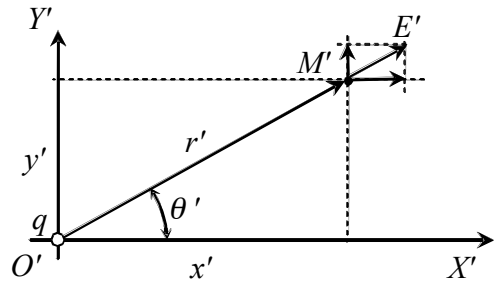

*Figure 1*. Scenario for deriving the electric and magnetic fields generated by a uniformly moving point charge in its rest frame K' (X'O'Y').

The first problem we have to solve is to find out a relationship between the field $E$ generated by the charge as measured by observers from K and the same field $E'$ measured by observers from K'. Thought experiments involving infinite extending charged surfaces or lines lead to[4]

$$E_x = E'_x \qquad (8)$$

and

$$E_y = \frac{E'_y}{\sqrt{1-\left(\frac{V}{c}\right)^2}} \qquad (9)$$

(6) and (7) becoming

$$E'_x = \frac{kqx'}{r'^3\sqrt{1-\left(\frac{V}{c^2}\right)}} \qquad (10)$$

$$E_y = \frac{kqy'}{r'^3\sqrt{1-\left(\frac{V}{c}\right)^2}} \qquad (11)$$

It is customary in special relativity to express the right sides of (10) and (11) as a function of physical quantities measured in K. Because the charge is a velocity independent physical quantity, a fact best proved by the neutrality of an atom inside which negative electrons move with different values around the positive nucleus, the only thing we have to do is to express $x', y', r'$ as a function of $x, y, r$ via the Lorentz-Einstein transformations.

There are two possible approaches. In the first one, observers from K detect simultaneously the two ends of the position vector of point $M'$, a procedure associated with the two simultaneous events $E_0(0,0,0)$ and $E(x, y, 0)$ resulting that

$$x' = \frac{r\cos\theta}{\sqrt{1-\left(\frac{V}{c}\right)^2}} \qquad (12)$$

$$y' = y = r\sin\theta \qquad (13)$$

$$r' = r\frac{\sqrt{1-\beta^2\sin^2\theta}}{\sqrt{1-\left(\frac{V}{c}\right)^2}} \qquad (14)$$

Equations (10) and (11) become

$$E_x = kq\frac{(1-\beta^2)\cos\theta}{r^2\left[1-\beta^2\sin^2\theta\right]^{3/2}} \qquad (15)$$

$$E_y = kq\frac{(1-\beta^2)}{}$$

or

$$\mathbf{E} = k(1-\beta^2)\frac{\mathbf{r}}{r^3(1-\beta^2\sin^2\theta)^{3/2}} \tag{16}$$

recovering (1).

In a second approach, which is more in the spirit of classical electromagnetism, we consider that the information about the fact that charge $q$ arrives at $t'=0$ at point O'(0,0) propagates in free space with speed $c$ and arrives at a point $M'(x'=r'\cos\theta', y'=r'\sin\theta')$ at a time $t'=\frac{r'}{c}$ generating the event $E'(x'=r'\cos\theta', y'=r'\sin\theta', t'=\frac{r'}{c})$. The event associated with the start of the information from O' is $E_0'(0,0,0)$. The same events detected from K are $E_0(0,0,0)$ and $E(x=r\cos\theta, y=r\sin\theta, t=\frac{r}{c})$. Appling the Lorentz-Einstein transformations we obtain

$$x' = r\frac{\cos\theta - \beta}{\sqrt{1-\left(\frac{V}{c}\right)^2}} \tag{17}$$

$$y' = y = r\sin\theta \tag{18}$$

$$r' = r\frac{1-\beta\cos\theta}{\sqrt{1-\left(\frac{V}{c}\right)^2}} \tag{19}$$

which with (10) and (11) become in this case

$$E_x = kq\frac{(\cos\theta - \beta)(1-\beta^2)}{r^2(1-\beta\cos\theta)^3} \tag{20}$$

$$E_y = kq\frac{(1-\beta^2)\sin\theta}{r^2(1-\beta\cos\theta)^3} \tag{21}$$

or

$$\mathbf{E} = \frac{kq}{r^3}\frac{(1-\beta^2)}{(1-\beta\cos\theta)^3}\left(\mathbf{r} - r\frac{\mathbf{V}}{c}\right). \tag{22}$$

By this simple deduction we recover Jefimenko's result[2]. However his original deduction appears to us as a veritable "tour-de-force" involving a great number of intermediary steps.

The magnetic field can be calculated using the well known formula that relates the electric and the magnetic field
$$\mathbf{B} = c^{-2}\mathbf{V} \times \mathbf{E}. \tag{23}$$
We consider that our second approach to derive the space coordinates of the point where the field is measured involving a radar detection procedure is more in the spirit of electrodynamics than the first approach involving the simultaneous detection of the moving rod coordinates[5]. We also point out that the radar detection shares much in common with the concept of retardation.

**Conclusions**

We have presented a new method to calculate the electric and magnetic fields generated by an electric charge moving with constant velocity. Our method is based on the radar detection of the point space coordinates where the fields are measured. This calculation is much simpler than an alternative approach[2] based exclusively on general electromagnetic equations which doesn't make use of retarded potentials or relativistic equations. In accordance with a hint of Ockham we propose the use of our simpler deduction in teaching relativistic electrodynamics as it is more transparent and time saving.